# Direct 3D information fusion for depth of field enhancement in optical-resolution photoacoustic microscopy

Xianlin Song [a, *], Sihang Li [b], Zhuangzhuang Wang [b]

[a] School of Information Engineering, Nanchang University, Nanchang 330031, China;

[b] Ji luan Academy, Nanchang University, Nanchang 330031, China;

[*] Corresponding author: songxianlin@ncu.edu.cn

## ABSTRACT

As an important branch of photoacoustic microscopy, optical-resolution photoacoustic microscopy suffers from limited depth of field due to the strongly focused laser beam. In this work, a 3D information fusion algorithm based on 3D stationary wavelet transform and joint weighted evaluation optimization is proposed to fuse multi-focus photoacoustic data to achieve large-volumetric and high-resolution 3D imaging. First, a three-dimensional stationary wavelet transform was performed on the multi-focus data to obtain eight wavelet coefficients. Differential evolution algorithm based on joint weighted evaluation was then employed to optimize the block size of division for each wavelet coefficient. Corresponding sub-coefficients of multi-focus 3D data were fused with the proposed fusion rule utilizing standard deviation for focus detection. Finally, photoacoustic microscopy with large depth of field can be achieved by applying the inverse stationary wavelet transform on the 8 fused sub-coefficients. The fusion result of multi-focus vertically tilted fiber shows that the depth of field of optical-resolution photoacoustic microscopy is doubled without sacrificing lateral resolution via the proposed method. Furthermore, the effectiveness of the proposed method was verified through the fusion results of multi-focus vessel data. Our work provides a feasible solution for achieving large-volumetric, high-resolution photoacoustic microscopy for further data analysis, processing and applications.

## 1. INTRODUCTION

Photoacoustic imaging, which combines the advantages of the high resolution of optics imaging and the significant penetration depth of acoustic imaging[1], has been widely used in tumor detection[2], vascular imaging[3], and brain imaging[4] in recent years. Focused laser beam or focused ultrasound probe is commonly employed in photoacoustic microscopy (PAM) to achieve better spatial resolution. Based on the lateral resolution of the system, PAM can be categorized as optical-resolution photoacoustic microscopy (OR-PAM)[5] and acoustic-resolution photoacoustic microscopy (AR-PAM)[6]. AR-PAM illuminates the sample with a weakly focused laser beam, and the lateral resolution of AR-PAM is determined by the size of the acoustic focus. OR-PAM takes advantage of a strongly focused laser beam by contrast, and the lateral resolution of OR-PAM depends on the size of the optical focus, which results in rapid degradation of lateral resolution of the objects positioning outside the focal plane. Acquisition of large-scale imaging with precise structural information is of vital importance for accurate pathology analysis in clinical diagnosis. Multiple sets of 3D photoacoustic data with different focuses can be acquired with the conventional OR-PAM by axial mechanical scanning[7], and the fusion of the multi-focus 3D photoacoustic data to obtain large-volumetric, high-resolution photoacoustic image without sacrificing spatial resolution is an important issue to be addressed.

Previous methods for extending depth of field (DoF) of photoacoustic imaging include improving the transducer of

PAM[8], employing Bessel beam[9] and introducing new illumination modes[10], etc. However, these methods enhance the DoF of imaging by improving system hardware, which are inconvenient to transfer and modify. In addition, hardware-based DoF expansion methods require high cost. In recent years, information fusion has shown great potential in expanding the DoF of microscopic imaging[11-12] and some research based on information fusion has been conducted in photoacoustic imaging. Zhao et al. demonstrated that the fusion of photoacoustic images obtained via different illumination schemes based on the designed system shows more details in comparison to imaging through a single illumination scheme[13]. However, this system is mainly designed for photoacoustic computed tomography, and hardware improvement for imaging system is also required. Cao et al. employed a 2D image fusion algorithm to fuse the B images of multi-focus data and generated PAM with large DoF[14]. However, this method does not achieve 3D information fusion for photoacoustic data.

In this work, a 3D fusion method of photoacoustic information is proposed to achieve large-volumetric and high-resolution photoacoustic imaging. Photoacoustic data with different focuses are transformed into stationary wavelet domain, respectively, and the differential evolution (DE) algorithm based on hybrid evaluations is introduced to search for optimal block size. The wavelet coefficients are then fused by a designed standard deviation (STD) guided fusion rule. Finally, all-in-focus photoacoustic data can be achieved by inverse stationary wavelet transform (ISWT). The fusion results of the multi-focus vertically tilted fiber demonstrate that the proposed method can increase the DoF of OR-PAM by a factor of 2 while maintaining lateral resolution. The effectiveness of the proposed method is validated through the fusion of a set of multi-focus vessel data.

## 2. METHODS

### 2.1 3D fusion based on joint weighted evaluation optimization

We propose a 3D photoacoustic information fusion algorithm combining stationary wavelet transform[15], STD-based focus detection and differential evolution optimization. The multi-focus 3D photoacoustic data was firstly decomposed into 8 wavelet coefficients via 3D stationary wavelet transform (3D-SWT), each of which contains unique frequency information. The coefficients were then divided into several equal-sized blocks and STD is used to detect whether a block locates within the depth of field automatically[16]. The blocks detected as focused were preserved in the fused coefficients thereafter and large-volumetric and high-resolution photoacoustic data can be achieved by inverse stationary wavelet transform. As the block size significantly influences the fusion, we updated the size of the blocks through DE algorithm to adaptively obtain the optimized block size. As shown in Figure 1, the multi-focus photoacoustic data $P_1$ and $P_2$ with size of H×W×L were obtained through the virtual OR-PAM[17]. $P_1$ and $P_2$ were decomposed into 8 wavelet coefficients LLL, LLH, LHL, LHH, HHL, HHH, HLH and HLL (where H represents high-frequency filtering, L represents low-frequency filtering) using 3D-SWT, respectively. Decomposed photoacoustic signal allows for proper processing of each sub-coefficient with unique frequency information. Compared with the multi-focus data in the spatial domain, the transformation is considered as feature extraction of raw data. Deeper features of photoacoustic signal can be extracted for subsequent analysis and processing. The wavelet coefficients were divided into several sub-blocks with equal size of h×w×l based on spatial location, and the fusion result varies with the block size. Oversized block may include both focused and defocused region and the fused data may contain out-of-focus area. Undersized block may not be able to provide sufficient data points for accurate focus detection. Therefore, DE algorithm was employed to iterate the size of the block region to obtain the optimized block size and corresponding fusion result adaptively[18]. A weighted evaluation function $\mathcal{L}$ is proposed to evaluate the maximum amplitude projection (MAP) image of fused 3D data which combines three quantification parameters, namely information entropy (EN)[19], average gradient (AVG)[20], and structural similarity index (SSIM)[21] as shown in formula (1)

$$\begin{aligned}\mathcal{L} &= \mathcal{L}(\mathcal{L}_{\text{AVG}}, \mathcal{L}_{\text{EN}}, \mathcal{L}_{\text{SSIM}}) \\ &= \lambda_{\text{AVG}}\mathcal{L}_{\text{AVG}} + \lambda_{\text{EN}}\mathcal{L}_{\text{EN}} + \lambda_{\text{SSIM}}\mathcal{L}_{\text{SSIM}}\end{aligned} \quad (1)$$

Where $\mathcal{L}_{AVG}$, $\mathcal{L}_{EN}$ and $\mathcal{L}_{SSIM}$ are the quantitative evaluations of the MAP image of the fused 3D data. $\lambda_{AVG}$, $\lambda_{EN}$ and $\lambda_{SSIM}$ are the pre-defined weights for the three quantification parameters. Limited DoF results in information loss and partial blur in OR-PAM. Therefore, the weight $\lambda_{SSIM}$ is set relatively smaller to avoid blur and information loss. $\lambda_{AVG}$ and $\lambda_{EN}$ are set larger to enhance texture details and increase the amount of information. The weights are set to $\lambda_{AVG} = 0.6$, $\lambda_{SSIM} = 0.1$ and $\lambda_{EN} = 0.3$ in the experiments. Larger $\mathcal{L}$ of the fused data shows a better fusion quality. To find the optimal block size of sub-coefficients, the optimization problem was solved through DE as shown in expression (2).

$$(\max_{h,w,l}) \, (\mathcal{L} = \lambda_{AVG}\mathcal{L}_{AVG} + \lambda_{EN}\mathcal{L}_{EN} + \lambda_{SSIM}\mathcal{L}_{SSIM}) \qquad (2)$$

Based on the optimized block size given by DE, the STD of corresponding sub-blocks of coefficients decomposed from multi-focus data are compared to detect the focus regions. The blocks with larger STD, which contain more details, are considered to be within focused regions and kept as sub-blocks of fused coefficients. Blocks with smaller STD, which are considered to be out of focused regions, are discarded. The fused 3D data with extended DoF can be achieved by performing ISWT on the fused coefficients. Figure 1 takes the wavelet coefficient LLL as an example to illustrate the whole process of focus detection and block size optimization. The other seven wavelet coefficients are processed in the same way.

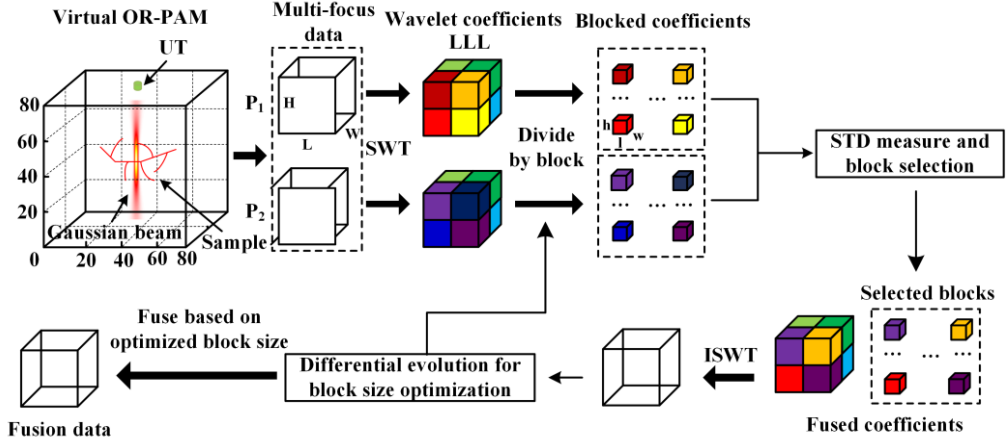

Figure 1. Flowchart of 3D photoacoustic information fusion based on joint weighted evaluation optimization. UT, ultrasonic transducer; P1, P2, multi-focus 3D photoacoustic data; SWT, stationary wavelet transform; ISWT, inverse stationary wavelet transform.

## 2.2 Multi-focus 3D data acquisition using virtual OR-PAM

A virtual Gaussian-beam photoacoustic microscopy based on k-Wave[16] was adopted to obtain multi-focus 3D photoacoustic data in our experiment. The Gaussian beam in the virtual OR-PAM platform is generated by an objective lens with a numerical aperture of NA = 0.2, and the light wavelength was set as 532 nm. The number of three-dimensional grids is Nx ×Ny×Nz = 80×80×80 and the pixel size is dx = dy = 2 μm, dz = 3 μm. Water was used as the medium around the sample, and the speed of sound was set to 1500 m/s. A ultrasonic detector with a center frequency of 75 MHz and a bandwidth of 67% was used to collect photoacoustic signals. The sample (such as vessel) was set up as required and placed in the 3D grid. Diagonally positioned fiber with a diameter of about 2 μm was imaged and fused to measure the performance of proposed 3D fusion algorithm. A set of simulated vessels was used to further validate the effectiveness of proposed method. The photoacoustic signal was obtained by raster scanning and reconstruction. 3D photoacoustic microscopy of different focal positions can be acquired by changing the position of the optical focus. All simulations were performed with a 64-bit Windows 10, Intel(R) Core(TM) i7-10750H CPU @ 2.60GHz desktop running windows operating system.

## 3. RESULTS

### 3.1 3D fusion based on joint weighted evaluation optimization

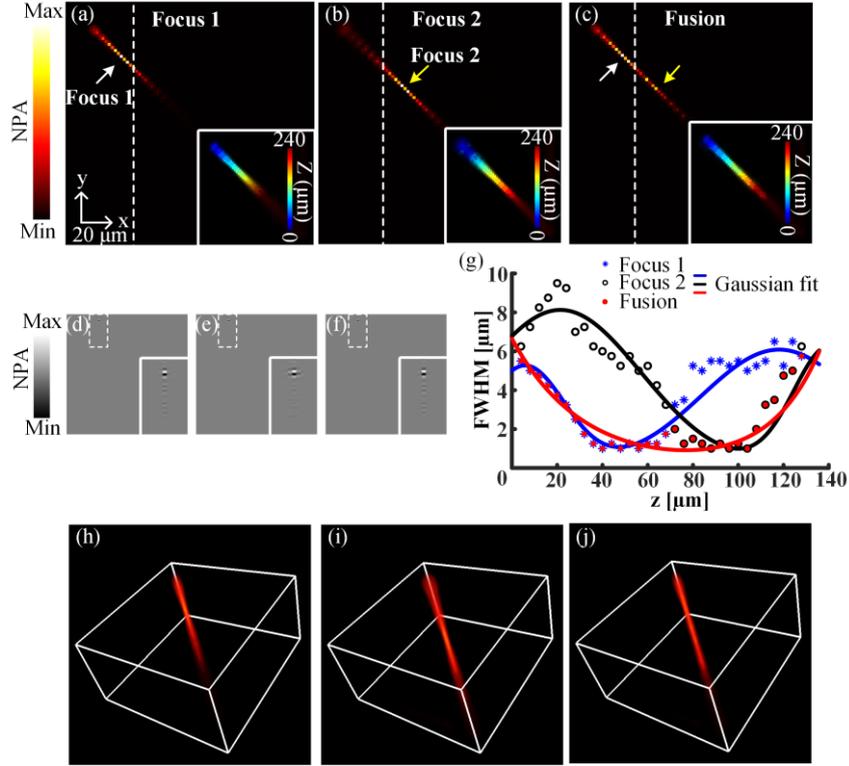

Figure 2. (a) and (b) are the MAP images projected from the tilted fibers obtained by the virtual OR-PAM when the focus positions are z = 30 and z = 40, respectively. (c) is the MAP image of the fusion data obtained through proposed method. Depth-coding MAP images of (a)-(c) are presented in the lower right corner, respectively. NPA is the normalized photoacoustic amplitude. (d)-(f) are the B images corresponding to the white dotted line in (a)-(c). (g) is the variation of lateral resolution along with the depth before and after fusion. (h)-(j) are the 3D visualizations rendered by Amira software corresponding to (a)-(c).

The multi-focus vertically tilted fibers and fusion results are shown in Figure 2. Figures 2(a) and 2(b) are a set of multi-focus vertically tilted fibers obtained by the virtual OR-PAM. Figure 2(a) is the MAP image of the tilted fiber with optical focus positioning at z = 30 in the 3D grid. Figure 2(a) is the MAP image of the tilted fiber with optical focus positioning at z = 40 (the optical focus is moved down by 30 μm) in the 3D grid and Figure 2(c) is the MAP image of fusion result of the tilted fibers through proposed method. The depth-coding MAP images of Figures 2(a)-2(c) are displayed in the lower right corner, repectively. The white arrows indicate the focal plane of Focus 1, and the yellow arrows indicate the focal plane of Focus 2. NPA is the normalized photoacoustic amplitude. It can be observed from Figures 2(a) and 2(b) that due to the limited DoF of OR-PAM, high-resolution and large-volumetric titled fiber cannot be captured during a single imaging. The lateral resolution at the focal plane is the best, and the imaged fiber located within the focal plane is the smallest in width. The lateral resolution and signal-to-noise ratio of imaging out of focal plane degrade rapidly, and the imaging becomes partly blurred. The clear portion of imaging varies with the positions of optical focus. The upper and lower halves of the fused fiber preserve the in-focus portion of source data and the fused image has a larger DoF, as shown in Figure 2(c). Figures 2(d)-2(f) are the B images corresponding to the dotted lines in Figures 2(a)-2(c). Figure 2(g) shows full width at half maximum (FWHM) of the profile of the fiber before and after fusion. Smaller FWHM represents a better lateral resolution. The black circle corresponds to Focus 1, the blue star corresponds to Focus 2, and the red circle corresponds to Fusion. The lateral resolution of the focused part (0-70 μm along z direction of Focus 1 and 70-120 along z direction μm of Focus 2) is small, while lateral resolution of the defocused part (70-120 μm along z direction of Focus 1 and 0-70 μm along z direction of Focus 2) is relatively large, as shown in Figure 2(c). The DoF of OR-PAM is defined as the depth range in which the FWHM of the fiber becomes twice the nadir. The DoF of Focus 1 is estimated to be about 37.3 μm where the FWHM varies from 1.1 μm to 2.2 μm. The DoF of Focus 2 is estimated to be about 33.0 μm where the FWHM varies from 0.99 μm to 1.98 μm. Through our proposed method, the DoF of OR-PAM is extended to 67.2 μm where FWHM varies from 0.91 μm to 1.82 μm. The DoF assessments of fibers before and after fusion quantitively show that the DoF of OR-PAM can be doubled without the sacrifice of lateral resolution through our proposed method. To observe the 3D structure of fibers more intuitively, the above three sets of photoacoustic data were imported into the Amira software for 3D visualization as shown in Figures 2(h)-2(j).

### 3.2 Large volumetric imaging of vessel

The effectiveness of the proposed method in processing sophisticated photoacoustic information is verified by fusing simulated vessels. The multi-focus vessel data and fusion results are presented in Figure 3. Figures 3(a) and 3(b) are a set of MAP images projected from multi-focus simulation vessels obtained by the virtual OR-PAM. The positions of optical focus set in Figures 3(a) and 3(b) are z = 40 and z = 60 (The optical focus is moved down by 60 μm). NPA is the normalized photoacoustic amplitude. The lower right corner of Figures 3(a)-3(c) are the close-up MAP images of the areas indicated by the white dashed rectangles in the figures, respectively. From Figures 3(a) and 3(b), it can be seen that due to the limited DoF of OR-PAM, the whole vessel cannot be captured through a single imaging. Figure 3(c) is the MAP image of the fusion results obtained using the proposed method. The upper and lower halves of the fused vessels are in-focus and the fusion result is all-in-focus imaging with larger DoF. The method proposed in this paper can still preserve the details of the multi-focus data when the imaged sample has a more sophisticated structure as shown in Figures 3(a)-3(c). To observe the detailed structure of the vessel, the above three sets of photoacoustic data were imported into the Amira for 3D visualization as shown in Figures 3(d)-3(f). One can observe that the imaging vessel (indicated by the white dashed rectangle in Figure 3(d)) located within the focused area shows a clear structure. The vessel located in the defocused area (indicated by the white dashed rectangle in Figure 3(e)) is blurred due to the limited DoF. With the method proposed, the fusion vessel (Figure 3(f)) retains the high resolution in Figure 3(e). Figures 3(g)-3(i) are the normalized amplitude distributions along the dashed lines in the partially enlarged views of Figures 3(a)-3(c) correspondingly. The FWHM of the signal intensity distribution within focused regions with a better lateral resolution shown in Figure 3(g) is measured to be 3.4 μm, while the FWHM of the signal intensity distribution positioning out of focused regions with a worse lateral resolution in Figure 3(g) is measured to be 9 μm. The corresponding signal intensity distribution of the fused vessel (Figure 3(i)) preserves the better lateral resolution of Figure 3(g), which further verifies the effectiveness of the proposed method in processing complex photoacoustic information.

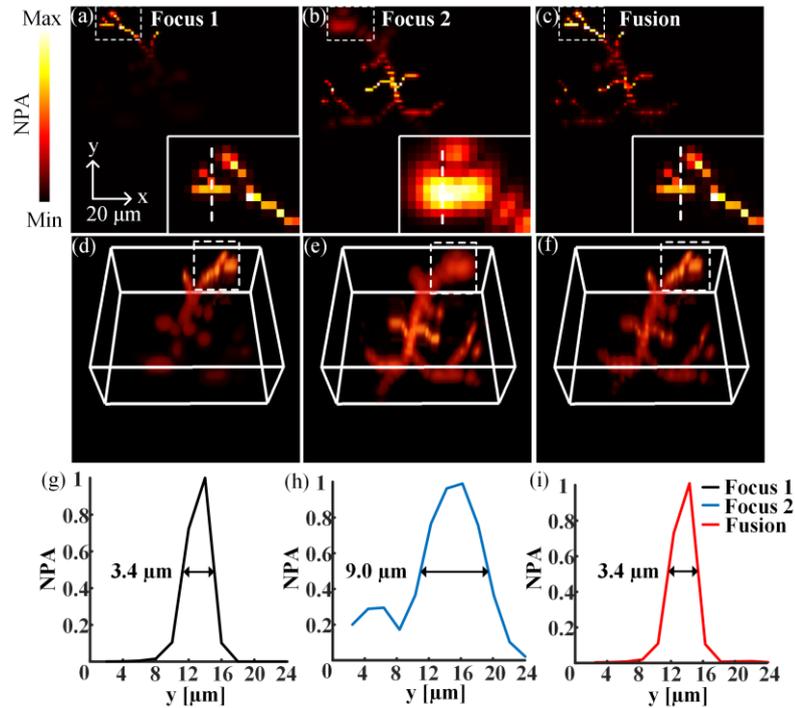

Figure 3. (a) and (b) are the MAP images projected from the simulated vessels obtained by the virtual OR-PAM when the focus positions are z = 40 and z = 60, respectively. (c) is the MAP image of the fusion data obtained through proposed method. The lower right corners of the (a)-(c) present the partially enlarged view of the white dotted rectangles in the figure. NPA is the normalized photoacoustic amplitude. (d)-(f) are the 3D visualizations rendered by Amira software corresponding to (a)-(c). (g)-(i) are the normalized intensity distributions of the white dotted lines in the enlarged images of (a)-(c).

## 4. CONCLUSION

A 3D multi-focus photoacoustic information fusion algorithm based on 3D-SWT and joint weighted evaluation optimization is proposed to extend the DoF of OR-PAM. The proposed algorithm adaptively fuses the information of multi-focus data to obtain single large volumetric and high-resolution photoacoustic microscopy. Experiment results show that the proposed algorithm achieves a superior depth of field expansion performance. The DoF of OR-PAM can be doubled at low cost without sacrificing lateral resolution through proposed algorithm. Our work provides a practical solution for obtaining large-volumetric high-resolution photoacoustic microscopy at low cost which is of great significance for biomedical research.